\documentclass[twocolumn,pra,showpacs,aps]{revtex4}

\usepackage{amsmath}
\usepackage{amsbsy}
\usepackage{amsthm}
\usepackage{amssymb}
\usepackage{latexsym}
\usepackage[dvips]{color,graphicx}

\begin{document}

\title{Quantum Monte Carlo study of confined fermions in 
one-dimensional optical lattices.}

\author{Marcos Rigol}
\affiliation{Institut f\"ur Theoretische Physik III, Universit\"at Stuttgart, 
Pfaffenwaldring 57, D-70550 Stuttgart, Germany.}
\author{Alejandro Muramatsu}
\affiliation{Institut f\"ur Theoretische Physik III, Universit\"at Stuttgart, 
Pfaffenwaldring 57, D-70550 Stuttgart, Germany.}

\begin{abstract}
Using quantum Monte Carlo (QMC) simulations we study the ground-state
properties of the one-dimensional fermionic Hubbard model in traps with 
an underlying lattice. Since due to the confining potential the density 
is space dependent, Mott-insulating 
domains always coexist with metallic regions, such 
that global quantities are not appropriate to describe the system. We 
define a local compressibility that characterizes the Mott-insulating regions
and analyze other local quantities. It is shown that the momentum
distribution function, a quantity that is commonly considered in experiments,
fails in giving a clear signal of the Mott-insulator transition. Furthermore, 
we analyze a mean-field approach to these systems and compare it with 
the numerically exact QMC results. Finally, we
determine a generic form for the phase diagram that allows us to 
predict the phases to be observed in the experiments.
\end{abstract}
\maketitle

\section{Introduction} 

The realization of Bose-Einstein condensation (BEC) of trapped
atomic gases \cite{anderson,bradley,davis} has generated in the
last years a huge amount of experimental and theoretical research 
\cite{bose,dalfovo}. BEC
was achieved by confining atoms in magnetic traps and lowering the
temperature of the system via the evaporative cooling technique.
In this technique, the hottest atoms are selectively removed from
the system and the remaining ones rethermalize via two-body collisions.
A common feature of these experiments is that the 
trapped gases are dilute; mean-field theory then provides a 
useful framework to study the role of the interaction between particles. 

Recently, a new feature has been added to the experiments: the 
magnetically trapped condensate is transferred into an optical lattice
generated by interfering laser beams \cite{greiner1}. With this new
experimental setup it is possible to access the strongly 
correlated regime. The superfluid-Mott-insulator
transition was studied \cite{greiner1} and other interesting physical
phenomena such as the collapse and revival of the condensate 
have been found \cite{greiner2}. The presence of the optical
lattice and the fact that the particles interact only via contact
interaction lead in a natural way to the Hubbard model as a paradigm 
for these systems. A theoretical work proposing such experiments
\cite{jaksch} and recent quantum Monte Carlo (QMC) simulations
\cite{batrouni,kaskurnikov} have investigated these systems beyond 
the mean-field approximation. It has been found that the incompressible 
Mott-insulating phase always coexists with compressible phases so that
a local order parameter \cite{batrouni} has to be defined to
characterize the system. Also, the phase diagram of trapped bosons 
was found to be more complex than the one in the
homogeneous system \cite{batrouni}.

The experimental realization of ultracold fermionic gases is more
difficult than the bosonic case, and has been achieved only
recently \cite{demarco,truscott,schreck,granade,hadzibabic,roati,ohara}.
Unlike bosons, single species fermions cannot be directly evaporatively 
cooled to very low temperatures because the s-wave scattering 
that could allow the gas to rethermalize during the evaporation is 
prohibited for identical fermions. This problem has been
overcome by simultaneously trapping and evaporatively cooling 
two-component Fermi gases \cite{demarco,granade}, and introducing 
mixed gases of bosons and fermions
in which bosons enable fermions to rethermalize through their
elastic interactions \cite{truscott,schreck,hadzibabic,roati}. More
recently, rapid forced evaporation employing a Feshbach resonance of 
two different spin states of the same fermionic atom 
has been used \cite{ohara}. 
Now that it is possible to go well below the degeneracy temperature 
in the experiments and superfluidity appears within reach \cite{ohara}, 
it is expected that the metal-Mott-insulator transition (MMIT) could 
also be realized for fermions on an optical lattice.

Motivated by this expectation we study the ground state of the 
one-dimensional (1D) fermionic Hubbard model with a harmonic trap 
and with repulsive contact interaction, using QMC simulations. Like in 
the bosonic case \cite{batrouni}, Mott-insulating domains appear over a 
continuous range of fillings and always coexist with compressible phases,
such that global quantities are not appropriate to characterize the system. 
Instead, we define a local-order parameter (a local compressibility) 
in order to characterize the local phases present in the system. 
We also analyze the generic features that are valid for any kind of 
confining potential, and not only for the harmonic one.

It is well known that the Hubbard model in the periodic case
displays a MMIT phase transition at half filling and at a finite value
of the on-site repulsive interaction $U$. (In the case of perfect
nesting the transition occurs at $U=0$.) The two routes to this
MMIT are the filling-controlled MMIT and the bandwidth-controlled 
MMIT \cite{imada}. In the confined case, it is possible to drive 
the transitions by changing the total filling of the trap, the on-site
repulsive interaction, or varying the curvature of the confining 
potential. In the following section, we study a number of local quantities as 
a function of the filling and the strength of the interaction. Although 
they signal the appearance of a Mott-insulating phase,
such quantities provide no rigorous criterion to characterize the 
Mott-insulating region. We therefore define a local compressibility
that acts as a genuine local-order parameter. 
In Sec.\ III, we study the momentum distribution function for the 
confined system. In Sec.\ IV, we discuss a mean-field approach for 
the 1D trapped system and compare the results with the ones obtained 
using QMC. In Sect.\ V, we analyze the phase diagram
and determine its generic form, which allows to compare systems 
with different sizes, number of particles and curvatures of the 
harmonic confining potential. In this section we also analyze the 
extension, for arbitrary confining potentials, of the results obtained 
for the harmonic case. Finally, the conclusions are given in 
Sec.\ VI. Some of the results presented here were summarized in a 
previous publication \cite{rigol1}.

\section{Local order parameter}

Due to the fact that the confining potential leads to an inhomogeneous 
density profile, we study in the present section how local quantities 
behave in the trapped system when the parameters at hand are changed. 
The Hamiltonian of the fermionic Hubbard model with a confining parabolic 
potential has the form
\begin{eqnarray}
H & = & -t \sum_{i,\sigma}( c^\dagger_{i\sigma} c^{}_{i+1
\sigma} + \textrm{H.c.}) + U \sum_i n_{i \uparrow} n_{i \downarrow}
\nonumber \\ & & + V \sum_{i \sigma} x_i^2\ n_{i \sigma}, \label{Hubb}
\end{eqnarray}
where $c^\dagger_{i\sigma}$, $c^{}_{i\sigma}$ are creation and 
annihilation operators, respectively, for a fermion with spin $\sigma$ 
at site $i$, and $n_{i \sigma} = c^\dagger_{i\sigma} c^{}_{i\sigma}$, 
such that $t$ is the hopping amplitude, $U$ is the on-site 
interaction that in the present work
will be considered repulsive ($U>0$), $V$ is the curvature of the
confining harmonic potential, and $x_i$ measures the position of the 
site $i$ ($x_i=ia$ with $a$ the lattice constant). The number of 
lattice sites is $N$ and is selected so that all the fermions are confined 
in the trap. We denote the total number of fermions in the trap as 
$N_f$ and consider equal number of 
fermions with spins up and down ($N_{f\uparrow}=N_{f\downarrow}=N_f/2$). 
In our simulations, we used the zero-temperature
projector method \cite{sugiyama,sorella} adapted from the QMC
determinantal algorithm by Blankenbecler, Scalapino, and Sugar
\cite{scalapino,blankenbecler,sugar}. The discrete
Hubbard-Stratonovich transformation by Hirsch was used \cite{hirsch}.
For details of the algorithm we refer to a number of reviews 
\cite{alejandro,fakher}.

\begin{figure}[h]
\includegraphics[width=0.46\textwidth,height=0.33\textwidth]
{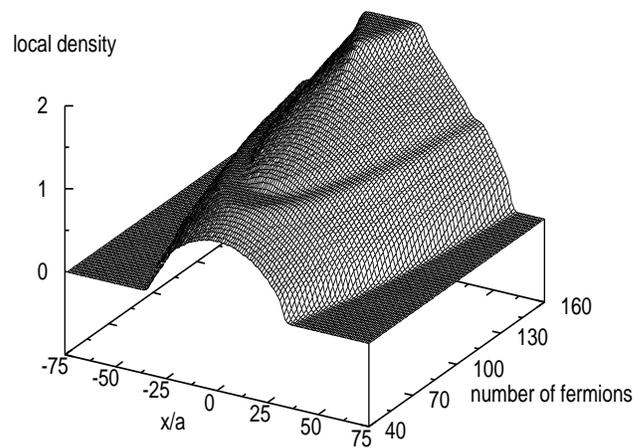}
\caption{Evolution of the local density in a parabolic confining 
potential as a function of the position in the trap and increasing total 
number of fermions. The parameters involved are $N=150$, $U=4t$ and 
$Va^2=0.002t$. The positions are measured in units of the 
lattice constant $a$.}
\label{perfil3D}
\end{figure}

The results for the evolution of the local density ($n_{i}=\langle
n_{i\uparrow }+n_{i\downarrow}  \rangle$), as a function of 
the total number of the confined particles, are shown in Fig.\ 
\ref{perfil3D}.
For the lowest filling, so that $n < 1$ at every site, the density shows 
a profile with the shape of an inverted parabola, similar to that 
obtained in the non-interacting case \cite{vignolo1}, and hence, such a 
situation should correspond to a metallic phase.
Increasing the number of fermions up to $N_f=60$, a plateau with
$n=1$ appears in the middle of the trap, surrounded by a region with
$n<1$ (metallic). Since in the homogeneous case, a Mott insulator appears 
at $n=1$, it is natural to identify the plateau with such a phase. 
The Mott-insulating domain in the center of the trap
increases its size when more particles are added, but at a certain
filling ($N_f=70$ here) this becomes energetically unfavorable and a new
metallic phase with $n > 1$ starts to develop in
the center of the system. Upon adding more fermions, this new metallic
phase widens spatially and the Mott-insulating domains surrounding it are
pushed to the borders. Depending on the on-site repulsion strength,
they can disappear and a complete metallic phase can appear in the
system. Finally, a ``band insulator'' (i.e., $n=2$) forms in the middle of 
the trap for the highest fillings (after $N_f=144$ here). Due to
the full occupancy of the sites, it will widen spatially and push the
other phases present in the system to the edges of the trap when
more fermions are added.

\begin{figure}[h]
\includegraphics[width=0.49\textwidth,height=0.41\textwidth]{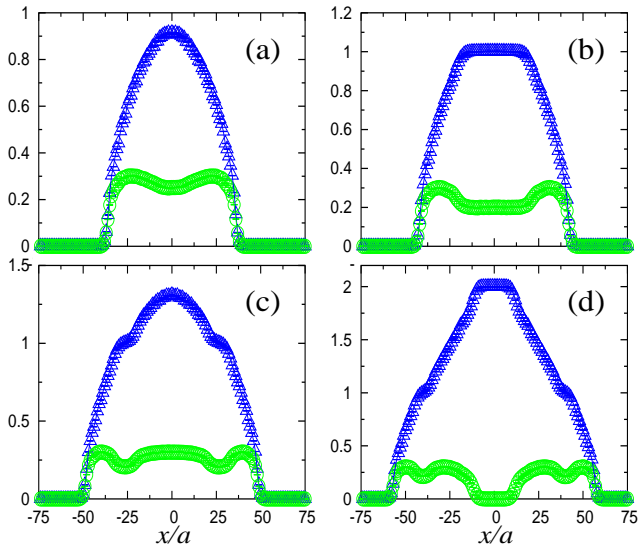}
\caption{(Color online) Four density profiles (\textcolor{blue}{$\triangle$}) 
(cuts across Fig.\ \ref{perfil3D}) and their variances 
(\textcolor{green}{$\bigcirc$}). The fillings are $N_f=$50 (a), 
68 (b), 94 (c), and 150 (d).}
\label{perfiles150}
\end{figure}

Although the existence of flat regions in the density profile 
is an indication that there is an insulator
there, a more quantitative characterization is needed. As shown in 
Ref.\ \cite{batrouni}, the variance of the
local density ($\Delta_i=\langle n^2_i \rangle - \langle n_i \rangle ^2$) 
may be used on a first approach (from here on, 
we refer to the variance as the variance of the local density).
In Fig.\ \ref{perfiles150} we show four characteristic profiles present 
in Fig.\ \ref{perfil3D} and their respective variances when the number 
of fermions in the system are $N_f=$50 (a), 68 (b), 94 (c), and 150 (d). 
For the case in which only a metallic phase is present [Fig.\
\ref{perfiles150}(a)], it is possible to see that the variance 
decreases when the density approaches $n=1$ and has a minimum for 
densities close to that value. For the Mott-insulating domain
[$n=1$ in Fig.\ \ref{perfiles150}(b)], the variance has a constant value
that is smaller than that of the metal surrounding it. As soon as the
Mott-insulating phase is destroyed in the middle of the trap and a
new metallic region with $n>1$ develops there [Fig.\ \ref{perfiles150}(c)], 
the variance increases in this region. The variance in the
metallic region with $n>1$ will start to decrease again when the
density approaches $n=2$ and will have values even smaller than
those in the Mott-insulating phase. Finally, when the insulator with $n=2$
is formed in the center of the trap the variance vanishes there 
[Fig.\ \ref{perfiles150}(d)]. 

\begin{figure}[h]
\includegraphics[width=0.43\textwidth,height=0.73\textwidth]
{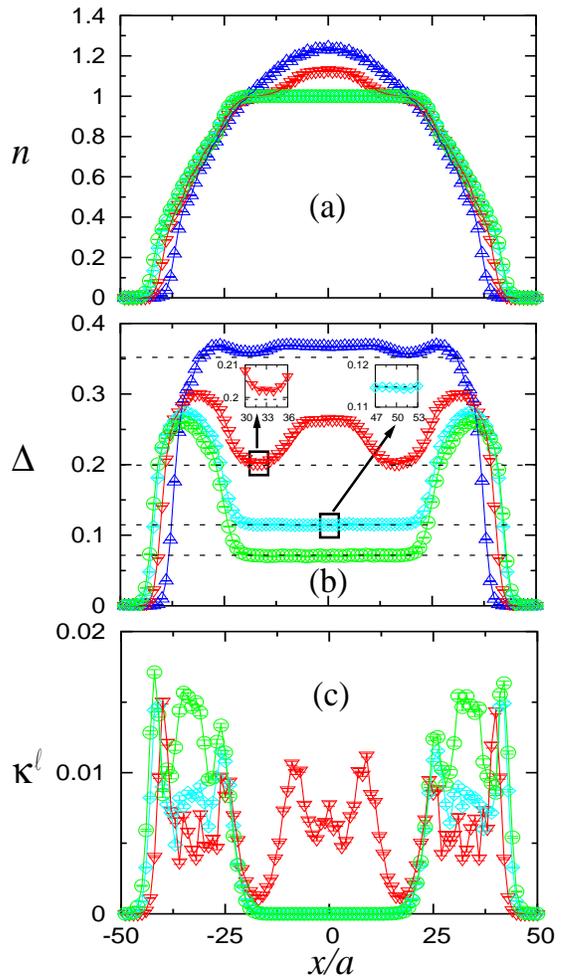}
\caption{(Color online) Profiles for a trap with $Va^2=0.0025t$ 
and $N_f=70$, the 
on-site repulsions are $U/t=$2 (\textcolor{blue}{$\triangle$}), 
4 (\textcolor{red}{$\bigtriangledown$}), 
6 (\textcolor{cyan}{\large $\Diamond$}), 
and 8 (\textcolor{green}{$\bigcirc$}). (a) Local density, 
(b) variance of the local density, (c) local compressibility 
$\kappa^\ell$ as defined in Eq.\ (\ref{localc}). 
The dashed lines in (b) are the values of the variance in the $n=1$ 
homogeneous system for $U/t=2, \ 4,\ 6,\ 8$ (from top to bottom).}
\label{perfdeltcomp100}
\end{figure}

Alternatively, Mott-insulating regions can be obtained 
by increasing the ratio between the on-site repulsive interaction 
and the hopping parameter as was done in experiments for bosons confined 
in optical lattices \cite{greiner1}.
Fig.\ \ref{perfdeltcomp100}(a) shows the evolution of the density
profiles in a trapped system with $N=100$, $N_f=70$, and
$Va^2=0.0025t$ when this ratio is increased from
$U/t=2$ to $U/t=8$ [for more details of these density profiles 
see Fig.\ \ref{PerfilK100}(e)-\ref{PerfilK100}(h)]. It can be seen that 
for small values of
$U/t$ ($U/t=2$) there is only a metallic phase present in the
trap. As the value of $U/t$ ($U/t=4$) is increased, a Mott-insulating phase tries to
develop at $n=1$ while a metallic phase with $n>1$ is present
in the center of the system. As the on-site repulsion is increased even 
further ($U/t=6,\ 8$), a Mott-insulating domain appears in the middle of the 
trap suppressing the metallic phase that was present there. In Fig.\ 
\ref{perfdeltcomp100}(b) we show the variance of the density for 
the profiles in Fig.\ \ref{perfdeltcomp100}(a) (from top to bottom, the 
values presented are for $U/t=2, \ 4, \ 6,\ 8$). As expected, the 
variance decreases in both the metallic and Mott-insulating phases when the on-site 
repulsion is increased. When the Mott-insulating plateau is formed in the density 
profile, a plateau with constant variance appears in the variance
profile with a value that will vanish only in the limit
$U/t \rightarrow \infty$. 
As shown in Fig.\ \ref{perfdeltcomp100}(b), whenever a Mott-insulating domain is
formed in the trap, the value of the variance in it is exactly the same
as the one for the Mott-insulating phase in the homogeneous system for the
same value of $U/t$ (horizontal dashed lines). This would support
the validity of the commonly used local density (Thomas-Fermi)
approximation \cite{butts97}. However, the insets in 
Fig.\ \ref{perfdeltcomp100}(b), show that this is not necessarily the 
case, since for $U/t=4$, the value of the variance in the Mott-insulating phase of 
the homogeneous system is still not reached in the trap, although the 
density reaches the value $n = 1$. Therefore, in contrast to the homogeneous
case, a Mott-insulating region is not determined by the filling only.  
In the cases of $U/t=6$ [inset in Fig.\ \ref{perfdeltcomp100}(b) for a 
closer look] and $U/t=8$, the value of the variance in the homogeneous 
system is reached and then we can say that Mott-insulating phases are formed there.

Although the variance gives a first indication for the formation of
a local Mott insulator, an ambiguity is still present, since
there are metallic regions with densities very close to $n=0$ and 
$n=2$, where the variance can have even smaller values than in 
the Mott-insulating phases. Therefore, an unambiguous quantity is still
needed to characterize the Mott-insulating regions.
We propose a local compressibility as a local-order parameter
to characterize the Mott-insulating regions, that is defined as
\begin{equation}
\label{localc} \kappa_i^\ell = \sum_{\mid j \mid \leq \, \ell (U)}
\chi_{i,i+j} \ ,
\end{equation}
where
\begin{equation}
\chi _{i,j}=\left\langle n_{i}n_{j}\right\rangle -\left\langle
n_{i} \right\rangle \left\langle n_{j}\right\rangle
\end{equation}
is the density-density correlation function and $\ell (U) \simeq b
\, \xi (U)$, with $\xi (U)$ the correlation length of $\chi_{i,j}$
in the unconfined system at half-filling for the given value of
$U$. As a consequence of the charge gap opened in the Mott-insulating 
phase at half filling in the homogeneous system, the
density-density correlations decay exponentially [$\chi_{\left( x
\right)} \propto \exp^{-x/\xi\left( U\right)}$] enabling 
$\xi (U)$ to be determined. The factor $b$ is chosen within a 
range where $\kappa^\ell$ becomes qualitatively insensitive to its 
precise value. Since there is some degree of arbitrariness in the 
selection of $b$, we will examine its role in the calculation of 
$\kappa^\ell$ here. 
\begin{figure}[h]
\includegraphics[width=0.42\textwidth,height=0.24\textwidth]
{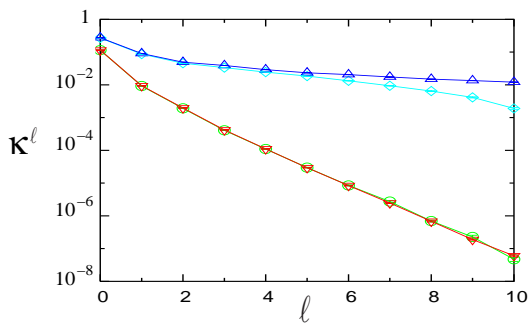}
\caption{(Color online) Local compressibility as a function of $\ell$ for 
the homogeneous and trapped ($Va^2=0.0025t$, $N_f=70$) systems in the 
Mott-insulating and metallic phases, 
$U=6t$. (\textcolor{red}{$\bigtriangledown$}) $n=1$ homogeneous,  
(\textcolor{green}{$\bigcirc$}) $n_{i=0}=1$ trapped, 
(\textcolor{blue}{$\triangle$}) $n = 0.66$ homogeneous, and 
(\textcolor{cyan}{\large $\Diamond$}) $n_{i=-32} \sim 0.66$ trapped. }
\label{Compvsl_QMC}
\end{figure}
We show in Fig.\ \ref{Compvsl_QMC} (in a semi-log plot) 
how the local compressibility behaves as a function of $\ell$ for a 
homogeneous and a trapped systems in the Mott-insulating ($n=1$) 
and metallic ($n\sim 0.66$) phases. 
In the Mott-insulating phase, it is possible to see 
that in the homogeneous and confined systems the local compressibility 
decays exponentially to zero in exactly the same way due to the charge 
gap present there. In the metallic case there is no charge gap so that 
density-density correlations decay as a power law and the local 
compressibility, shown in Fig.\ \ref{Compvsl_QMC}, decays very slowly. 
In Fig.\ \ref{Compvsl_QMC}, a departure in the behavior of the local 
compressibility for the trapped system from the homogeneous case for 
large values of $\ell$ can be seen. This occurs because 
correlations with points very close to the band insulating ($n=0$) and 
the Mott-insulating ($n=1$) regions surrounding the metallic phase [see 
Fig.\ \ref{perfdeltcomp100}(a)] are included in $\kappa$. 
However, this effect does not affect the fact that the value of the local 
compressibility is clearly different from zero as long as the size of the 
metallic phase is larger than $2\ell$. We obtain that considering 
$b\sim 10$ (no matter what the exact value of $b$ is), the local 
compressibility is zero with exponential accuracy (i.e.\ $\sim 10^{-6}$ 
or less) for the Mott-insulating phase and has a finite value in the 
metallic phase. For the case in Fig.\ \ref{Compvsl_QMC} where $U=6t$, the 
correlation length is $\xi \sim 0.8$ so that $\ell \sim 8$. 
Physically, the local compressibility defined here 
gives a measure of the change in the local density 
due to a constant shift of the potential over a finite range 
but over distances larger than the correlation length in the unconfined 
system.
 
In Fig.\ \ref{perfdeltcomp100}(c), we show the profiles of the local 
compressibility for the same parameters as Figs.\
\ref{perfdeltcomp100}(a) and (b) (we did not include the profile of the 
local compressibility for $U=2t$ because for that value of $U$ we obtain 
that $\ell$ is bigger than the system size). In Fig.\ 
\ref{perfdeltcomp100}(c), it can be seen that the local compressibility 
only vanishes in the Mott-insulating domains. For $U=4t$, it can be seen
that in the region with $n \sim 1$ the local compressibility,
although small, does not vanish. This is compatible with the fact that 
the variance is not equal to the value in the homogeneous 
system there, so that although there is a shoulder in the density profile, 
this region is not a Mott insulator. Therefore, the local compressibility 
defined here serves as a genuine local order parameter to
characterize the insulating regions that always coexist with
metallic phases. At this point we would like to remark that, as shown
in Ref.\ \cite{rigol1}, the local compressibility shows critical behavior
on entering the Mott-insulating region. This is the reason to speak about 
phases, since on passing from one region to the other, critical behavior 
sets in.

\begin{figure}[h]
\includegraphics[width=0.48\textwidth,height=0.44\textwidth]
{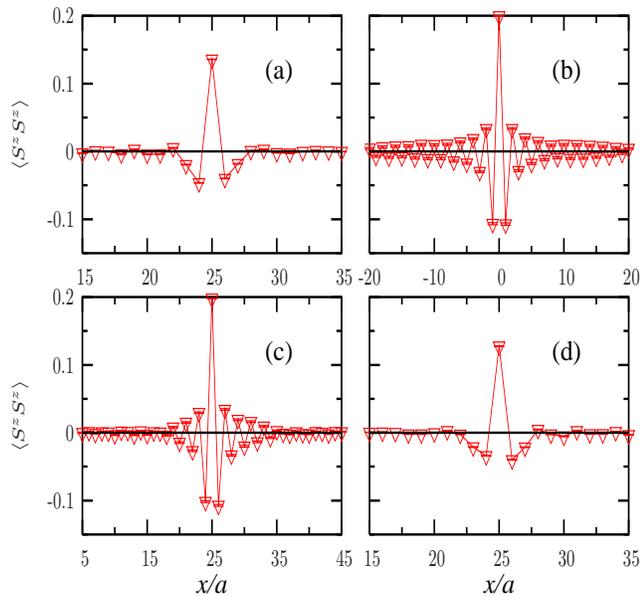}
\caption{(Color online) Local spin-spin correlations in a trap measured with 
respect to the points in which in the figure they have their maximum 
value. The density at each point (total filling in the trap) is 
$n_{i=25}=0.61$ ($N_f=$50) (a), $n_{i=0}=1.00$ ($N_f=$68) (b), 
$n_{i=25}=1.01$ ($N_f=$94) (c), and $n_{i=25}=1.44$ ($N_f=$150) (d), 
for a trap with $N=150$, $U=4t$, and $Va^2=0.002t$ 
(the density and variance profiles for these parameters 
were presented in Fig.\ \ref{perfiles150}).}
\label{Spinesz}
\end{figure}

Finally, we discuss in this section the spin-spin correlation function,
since in periodic chains, quasi-long-range antiferromagnetic 
correlations appear in the Mott-insulating phase. In Fig.\ \ref{Spinesz} 
we show the local $\langle S^z_i S^z_j \rangle$ correlation function for 
some points of the profiles presented in Fig.\ \ref{perfiles150}. 
We measured the spin-spin correlations in the trap at the points 
in the figures where it can be clearly seen that $\langle S^z_i S^z_j 
\rangle$ has the maximum value. In Fig.\ \ref{Spinesz}(a) it can be seen 
that in the metallic phase with $n<1$, the spin-spin correlations decay 
rapidly and do not show any clear modulation, which is due to  
the fact that the density is changing in this region. In the local 
Mott-insulating phase [center of the trap in Fig.\ \ref{Spinesz}(b)], 
short-range antiferromagnetic correlations are present, and they
disappear completely only on entering in the metallic regions. For the
shoulders with $n \sim 1$  [Fig.\ \ref{Spinesz}(c)], the 
antiferromagnetic correlations are still present but due to the small 
size of these regions they decay very rapidly. Finally for the metallic 
regions with $n>1$, the spin correlations behave like in the metallic 
phases with $n<1$ as shown in Fig.\ \ref{Spinesz}(d). 

\section{Momentum distribution function}

In most experiments with quantum gases carried out so far,
the momentum distribution function, which is determined
in time-of-flight measurements, played a central role. A
prominent example is given by the study of the 
superfluid-Mott-insulator transition \cite{greiner1} in the bosonic case.
Also a QMC study relating this quantity to the 
density profiles and proposing how to determine the point at which 
the superfluid-Mott-insulator transition occurs was presented in 
Ref.\ \cite{kaskurnikov}. As shown below, we find that this quantity is 
not appropriate to characterize the phases of the system in the fermionic case,
and does not show any clear signature of the MMIT.

\begin{figure}[h]
\begin{center}
\includegraphics[width=0.48\textwidth,height=0.44\textwidth]
{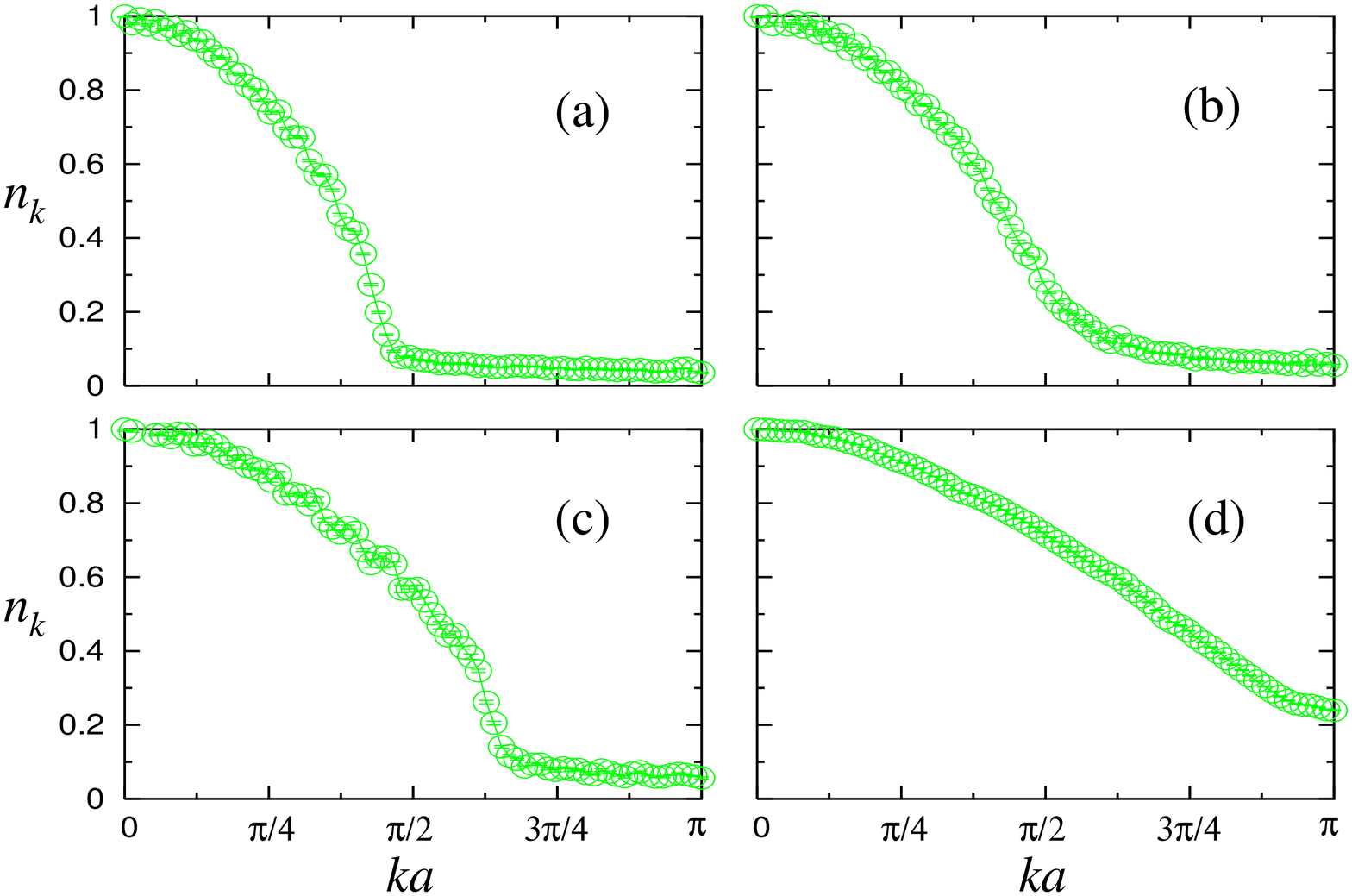}
\caption{(Color online) Normalized momentum distribution function 
for the same 
parameters of Fig.\ \ref{perfiles150}. The fillings are $N_f=$50 (a), 
68 (b), 94 (c), and 150 (d).}
\label{PerfilK150}
\end{center}
\end{figure}

In Fig.\ \ref{PerfilK150} we show the normalized momentum distribution 
function ($n_k$) for the same density profiles presented in 
Fig.\ \ref{perfiles150}. For the trapped systems, we always normalize the 
momentum distribution to be unity at $k=0$. In the case presented in 
Fig.\ \ref{PerfilK150}, transitions between phases occur due to changes 
in the total filling of the trap. We first notice that $n_k$ for the pure
metallic phase in the harmonic trap [Fig.\ \ref{PerfilK150}(a)] does not
display any sharp feature corresponding to a Fermi surface, in clear 
contrast to the homogeneous case. The lack of a sharp feature for the 
Fermi surface is independent of the presence of the interaction and is 
also independent of the size of the system. In the non-interacting case,
this can be easily understood: the spatial density and the
momentum distribution will have the same functional form because
the Hamiltonian is quadratic in both coordinate and momentum. When
the interaction is present, it could be expected that the formation of
local Mott-insulating domains generates a qualitatively and quantitatively
different behavior of the momentum distribution, like in the
homogeneous case where in the Mott-insulating phase the Fermi
surface disappears and $n_k$ is smoother. In Fig.\ \ref{PerfilK150}(b), 
it can be seen that there is no qualitative change of the momentum
distribution when the Mott-insulating phase is present in the
middle of the trap. Quantitatively $n_k$ in this case is very
similar to the pure metallic cases Fig.\ \ref{PerfilK150}(a) (for 
densities smaller than one) and Fig.\ \ref{PerfilK150}(c) 
(where in the middle of the trap the
density is higher than one). Only when the insulator with $n=2$ 
appears in the middle of the system we find a quantitative change 
of $n_k$ for $U=4t$, as shown in Fig.\ \ref{PerfilK150}(d).

\begin{figure}[h]
\begin{center}
\includegraphics[width=0.49\textwidth,height=0.74\textwidth]
{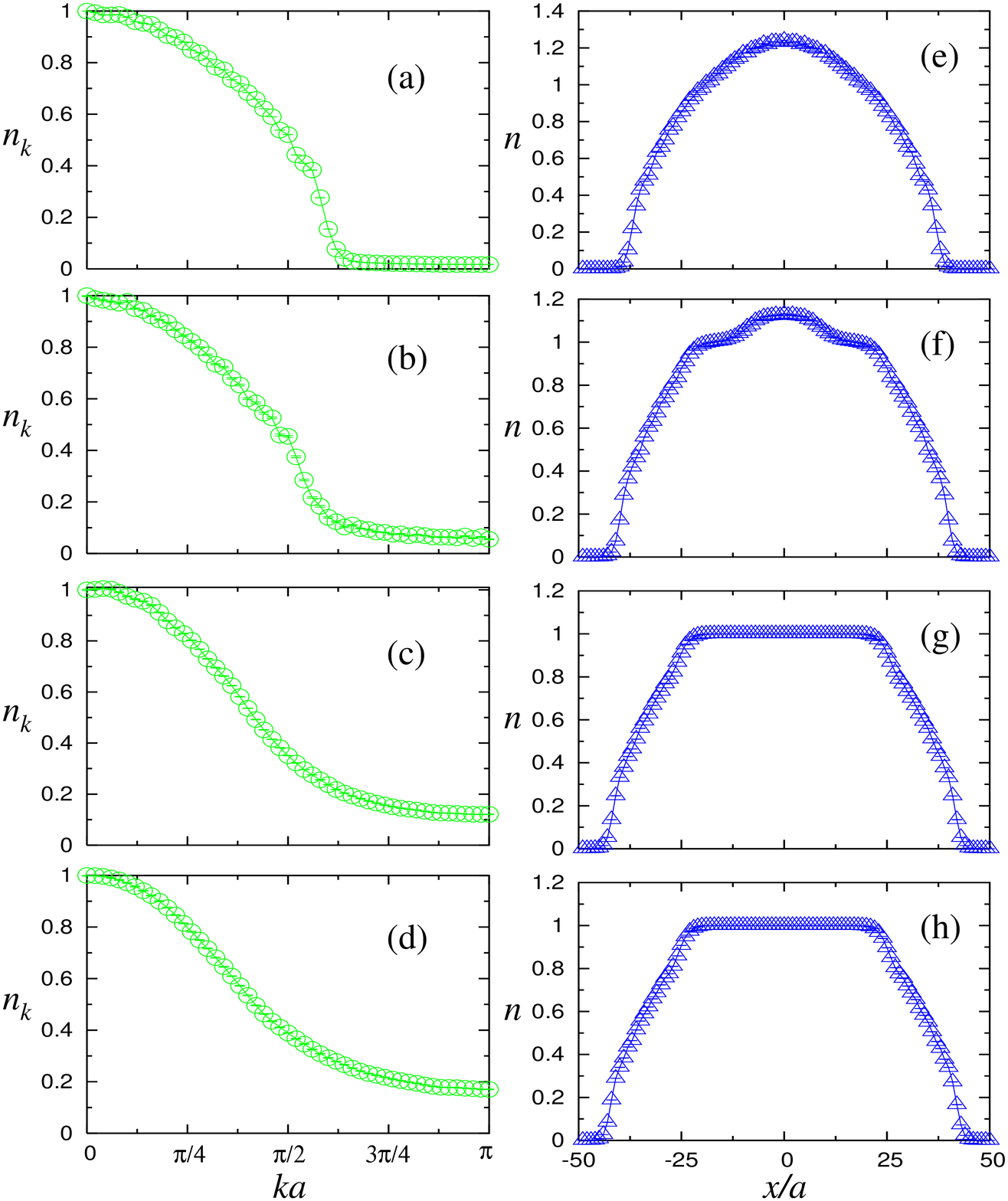}
\caption{(Color online) Normalized momentum distribution function 
(a)-(d) and their corresponding density profiles (e)-(h) for 
$U/t=2$ (a),(e), 4 (b),(f), 
6 (c),(g), 8 (d),(h) and $N=100$, $N_f=70$, $Va^2=0.0025t$.}
\label{PerfilK100}
\end{center}
\end{figure}

We also studied the momentum distribution function when the
MMIT is driven by the change of the on-site repulsion. We did not observe 
any clear signature of the formation of the Mott-insulating phase in 
$n_k$. In Fig.\ \ref{PerfilK100} we show the normalized momentum 
distribution function [Fig.\ \ref{PerfilK100}(a)-\ref{PerfilK100}(d)] 
for density profiles [Fig.\ \ref{PerfilK100}(e)-\ref{PerfilK100}(h)] 
in which the on-site repulsion was increased from $U/t=2$ to 8. It can be 
seen that the same behavior present in Fig.\ \ref{PerfilK150} and the 
quantitative changes in $n_k$ appear only when the on-site repulsion goes 
to the strong-coupling regime, but this is long after the Mott-insulating 
phase has appeared in the system.

At this point one might think that in order to study the MMIT using 
the momentum distribution function, it is necessary to avoid the 
inhomogeneous trapping potential and use instead a kind of magnetic box 
with infinitely high potential on the boundaries. However, in that case 
one of the most important achievements of the inhomogeneous system is 
lost, i.e., the possibility of creating Mott-insulating phases for a 
continuous range of fillings. In the perfect magnetic box, the 
Mott-insulating phase would only be possible at half filling, which would be 
extremely difficult (if possible at all) to adjust experimentally. 
The other possibility is to create traps 
which are almost homogenous in the middle and which have an appreciable 
trapping potential only close to the boundaries. This can be
studied theoretically by considering traps with higher powers of the
trapping potentials. As shown below, already non-interacting systems make 
clear that a sharp Fermi edge is missing in confined systems.

\begin{figure}[h]
\begin{center}
\includegraphics[width=0.4\textwidth,height=0.47\textwidth]
{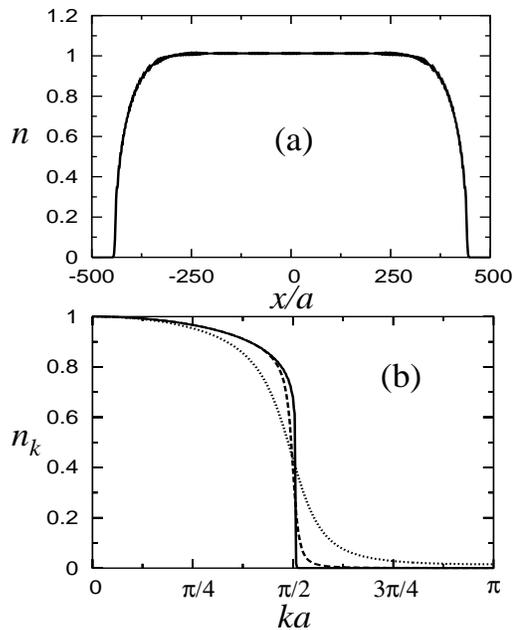}
\caption{Exact results for $N_f=840$ noninteracting trapped fermions in 
a lattice with 1000 sites and a confining potential 
$V_{10}a^{10}=7 \times 10^{-27}t$. Density profile (a) and the normalized 
momentum distribution function (b): the continuous line corresponds to 
(a), the dashed line is the result when an alternating potential 
$V_a=0.1t$ is superposed on the system, and the dotted line corresponds 
to $V_a=0.5t$.}
\label{PerfilK1000}
\end{center}
\end{figure}

In Fig.\ \ref{PerfilK1000}(a) we show the density 
profile of a system with 1000 sites, $N_f=840$, and a trapping 
potential of the form $V_{10}x_i^{10}$ with 
$V_{10}a^{10}=7 \times 10^{-27}t$. It can be seen that the density 
is almost flat all over the trap with a density of the order 
of one particle per site. Only a small part of the system
at the borders has the variation of the density required for the
particles to be trapped. In Fig.\ \ref{PerfilK1000}(b) (continuous line), 
we show the corresponding normalized momentum distribution. It 
can be seen that a kind of Fermi surface develops in the system but 
for smaller values of $k$, $n_k$ is always smooth and its value starts 
decreasing at $k=0$.
In order to see how $n_k$ changes when an incompressible region
appears in the system, we introduced an additional alternating
potential, so that in this case the new Hamiltonian has the form
\begin{eqnarray}
H & = & -t \sum_{i,\sigma} ( c^\dagger_{i\sigma} c^{}_{i+1
\sigma} + \textrm{H.c.} )  + V_{10} \sum_{i \sigma} x_i^{10}\ n_{i \sigma} 
\nonumber \\ & & + V_a  \sum_{i \sigma} \left(-1\right)^i 
n_{i \sigma} , \label{HubbIon}
\end{eqnarray}
where $V_a$ is the strength of the alternating potential. For the 
parameters presented in Fig.\ \ref{PerfilK1000}(a), we
obtain that a small value of $V_a$ ($V_a=0.1t$) generates a band insulator 
in the trap, which extends over the region with $n \sim 1$ (when $V_a=0$). 
However, the formation of this band insulator is barely reflected 
in $n_k$, as can be seen in Fig.\ \ref{PerfilK1000}(b) (dashed line). 
Only when the value of $V_a$ is increased and the system departs from the 
phase transition [$V_a=0.5t$, dotted line in Fig.\ \ref{PerfilK1000}(c)], 
does a quantitatively appreciable change in $n_k$ appear.

\begin{figure}[h]
\begin{center}
\includegraphics[width=0.4\textwidth,height=0.49\textwidth]
{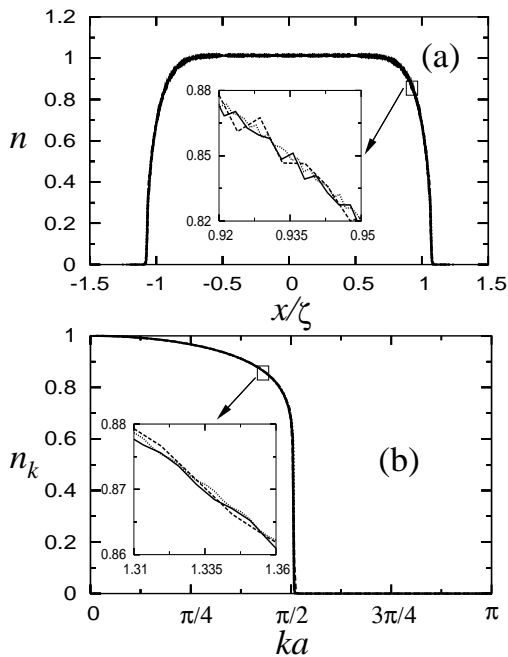}
\caption{Exact results for noninteracting trapped fermions in a 
confining potential $V_{10}$. (a) Density profiles, (b) normalized 
momentum distribution function. Dashed line corresponds to 
$V_{10}a^{10}=6 \times 10^{-24}t$, $N_f=428$ ($N \sim 500$), 
continuous line corresponds to $V_{10}a^{10}=7 \times 10^{-27}t$, 
$N_f=840$ ($N \sim 1000$) and dotted line corresponds to 
$V_{10}a^{10}=1 \times 10^{-29}t$, $N_f=1620$ ($N \sim 2000$). 
In the density profiles the positions are given in units of 
the characteristic length $\zeta$.}
\label{PerfilFree}
\end{center}
\end{figure}

In general, it is expected that on increasing the system size,
the situation in the homogeneous system is approached. However,
in the present case it is not merely a question of boundary conditions,
but the whole system is inhomogeneous. Therefore, a proper scaling has to 
be defined in order to relate systems of different sizes. In the case 
of particles trapped in optical lattices when there is a confining 
potential with a power $\alpha$ and strength ($V_{\alpha}$), 
a characteristic length of the system 
($\zeta$) is given by $\zeta=(V_{\alpha}/t)^{-1/\alpha}$, so 
that a characteristic density ($\tilde{\rho}$) can be defined as 
$\tilde{\rho}=N_f a/ \zeta$. We find that this characteristic 
density is the one meaningful in the thermodynamic limit. 
In Fig.\ \ref{PerfilFree} we show three systems in which the total number 
of particles and the curvature of the confining potential $V_{10}$ were 
changed, keeping $\tilde{\rho}$ constant. 
We measured the positions in 
the trap in units of the characteristic length $\zeta$. 
These systems have occupied regions ($n>0$) with very different sizes, 
of the order of 500 lattice sites for the dashed line, of the order of 
1000 sites for the continuous line, and of the order of 2000 lattice 
sites for the dotted line. Fig.\ \ref{PerfilFree}(a) shows that the 
density profiles scale perfectly when the curvature of the confining 
potential and the number of fermions are changed in the system, so that 
in order to show the changes in the density profile, we introduced an 
inset that expands a region around a density 0.85. In the case of the 
momentum distribution function [Fig.\ \ref{PerfilFree}(b)], it is 
possible to see that this quantity also scales very well so that almost 
no changes occur in the momentum distribution function when the occupied 
system size is increased, implying that in the thermodynamic limit the 
behavior of $n_k$ is different from the one in the homogeneous system. 
An expanded view of the region with $n_k$ around 0.87 is introduced to 
better see the scale of the differences in this region. In Sec.\ V we 
show that the characteristic density defined here is also the 
meaningful quantity to define the phase diagram of the system. 

\section{Mean-field approximation}

The mean-field (MF) approach has been very useful in the study of
dilute bosonic gases confined in harmonic potentials
\cite{dalfovo}. This theory for the order parameter associated
with the condensate of weakly interacting bosons for
inhomogeneous systems takes the form of the Gross-Pitaevskii
theory. When the bosonic condensate is loaded in an optical
lattice, it is possible to go beyond the weakly interacting 
regime and reach the strongly correlated limit for which a 
superfluid-Mott-insulator transition occurs. Also in this limit a MF 
study was done \cite{jaksch}, and the results were compared with exact
diagonalization results for very small systems, reporting a
qualitatively good agreement between both methods \cite{jaksch}.

In this section, we compare
a MF approximation with QMC results for the system under consideration.
It will be shown that the MF approach not only violates the 
Mermin-Wagner theorem, leading to long-range antiferromagnetic order
in one dimension, as expected, but also introduces spurious
structures in the density profiles.
In order to obtain the MF Hamiltonian, we 
rewrite the Hubbard Hamiltonian in Eq.\ (\ref{Hubb}) in the following 
form (up to a constant shift in the chemical potential):
\begin{eqnarray}
H &=&-t\sum_{i,\sigma }( c_{i\sigma }^{\dagger }c_{i+1\sigma
}+\textrm{H.c.}) +V\sum_{i,\sigma }x_i^2\
\hat{n}_{i\sigma}  \notag \\
&&+\frac{\left( 1-\Lambda \right) }{2}[ \left\langle \hat{n}
_{i}\right\rangle \hat{n}_{i}-\left\langle
\hat{n}_{i}\right\rangle ^{2}+
\widehat{\delta }_{n_{i}}^{2}]   \notag \\
&&-\frac{\Lambda }{2}[ \left\langle \widehat{\mu
}_{i}\right\rangle \widehat{\mu }_{i}-\left\langle \widehat{\mu
}_{i}\right\rangle ^{2}+ \widehat{\delta }_{\mu _{i}}^{2}],
\label{MF1}
\end{eqnarray}
where the density and the magnetization in each site are denoted
by $\hat{n}_{i} =\hat{n}_{i\uparrow }+\hat{n}_{i\downarrow }$
and $\hat{\mu}_{i} =\hat{n}_{i\uparrow }-\hat{n}_{i\downarrow }$,
respectively. The fluctuations of the density and magnetization
are given by $\widehat{\delta }_{n_{i}}
=\hat{n}_{i}-\left\langle \hat{n}_{i} \right\rangle$ and
$\widehat{\delta }_{\mu _{i}} =\hat{\mu}_{i}-\left\langle
\hat{\mu} _{i}\right\rangle$, respectively, and $\Lambda$ is an
arbitrary parameter that was introduced in order to allow for the most 
general variation in parameter space. The following relations for fermionic
operators were used:
\begin{eqnarray}
\hat{n}_{i\uparrow }\hat{n}_{i\downarrow }
&=&\frac{1}{2}\hat{n}_{i}^{2}-
\frac{1}{2}\hat{n}_{i}  \notag \\
\hat{n}_{i\uparrow }\hat{n}_{i\downarrow }
&=&-\frac{1}{2}\hat{\mu}_{i}^{2}+ \frac{1}{2}\hat{n}_{i}.
\end{eqnarray}

Neglecting the terms containing the square of the density and 
magnetization fluctuations in Eq.\ (\ref{MF1}), the following MF 
Hamiltonian
\begin{eqnarray}
H_{MF} &=&-t\sum_{i,\sigma }( c_{i\sigma }^{\dagger }c_{i+1\sigma
}+\textrm{H.c.}) +V\sum_{i,\sigma }x_i^2\
\hat{n}_{i\sigma
}  \notag \\
&&+\frac{1}{2}[ \left( 1-\Lambda \right) \left\langle \hat{n}
_{i}\right\rangle -\Lambda \left\langle \widehat{\mu
}_{i}\right\rangle
] \hat{n}_{i\uparrow }  \notag \\
&&+\frac{1}{2}[ \left( 1-\Lambda \right) \left\langle \hat{n}
_{i}\right\rangle +\Lambda \left\langle \widehat{\mu
}_{i}\right\rangle
] \hat{n}_{i\downarrow }  \notag \\
&&-\frac{\left( 1-\Lambda \right) }{2}\left\langle
\hat{n}_{i}\right\rangle ^{2}+\frac{\Lambda }{2}\left\langle
\widehat{\mu }_{i}\right\rangle ^{2} \label{MF2}
\end{eqnarray}
is obtained. Due to the inhomogeneity of the trapped system, an 
unrestricted Hartree-Fock scheme is used to determine the local densities 
$ \left\langle \hat{n}_{i}\right\rangle$ and local magnetic moments
$ \left\langle \widehat{\mu}_{i}\right\rangle$. Given a MF ground state 
$|\Psi_{MF}\rangle$, the minimum energy $E= \langle \Psi_{MF}|
 H |\Psi_{MF} \rangle$ (not the MF one $E_{MF}=\langle \Psi_{MF}|
 H_{MF} |\Psi_{MF}\rangle$) is
reached when $\Lambda=0.5$, so that all the results that follow
were obtained for this value of $\Lambda$.

We first recall some of the discrepancies between MF
approximations and the known valid facts
for the 1D homogenous system ($V=0$) as follows.

\begin{figure}[h]
\includegraphics[width=0.40\textwidth,height=0.21\textwidth]{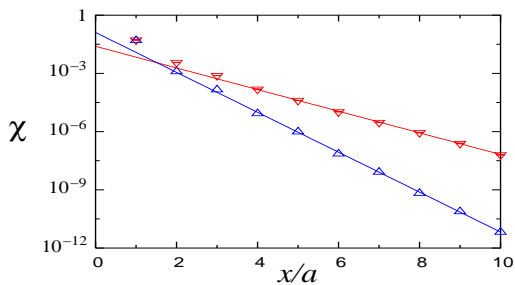}
\caption{(Color online) Density-density correlations for a homogeneous 
system with 
$N=102$ at half filling for $U=6t$. MF (\textcolor{blue}{$\triangle$}) 
and QMC (\textcolor{red}{$\bigtriangledown$}) solutions.}
\label{correlHomF}
\end{figure}

(i) At half filling it is known that the Hubbard model exhibits a 
Mott-insulating phase with a charge gap while the spin sector remains gapless, 
so that the density-density correlations decay exponentially and the 
spin-spin correlations decay as a power law. Within the MF approximation, 
there is a band insulator 
in the system at half filling so that both the density-density and 
spin-spin correlations decay exponentially. The MF value of the
charge gap is an overestimation of the real one, as can be seen in
Fig.\ \ref{correlHomF} where we present the MF results for the
density-density correlations as function of the distance, together
with the QMC results. 
The slopes of the curves are proportional to
the charge gap and the MF slope is approximately twice the QMC one (the
results were obtained for a system with 102 sites and $U=6t$).
The correlation length of $\chi$ is the inverse of these slopes. 
Finally, at half filling the MF theory leads to an antiferromagnetic 
state, while in the Hubbard model the magnetization is always zero
although quasi-long-range antiferromagnetic correlations appear in the
system.

\begin{figure}[h]
\includegraphics[width=0.39\textwidth,height=0.23\textwidth]
{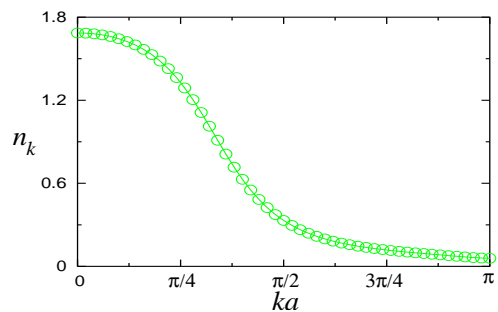}
\caption{(Color online) MF result for the momentum distribution function of 
a homogeneous system with $N=102$, $N_f=66$, and $U=6t$.}
\label{perfilK102o066MF}
\end{figure}

\begin{figure}[h]
\includegraphics[width=0.49\textwidth,height=0.24\textwidth]
{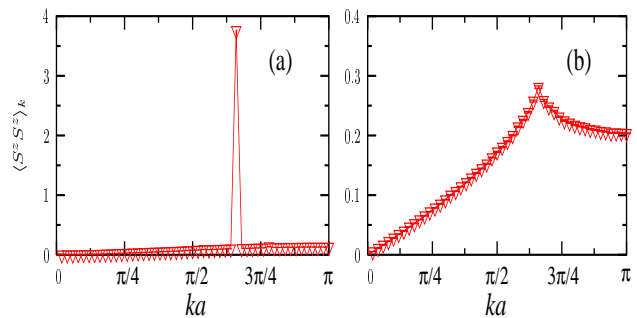}
\vspace{-0.7cm}
\caption{(Color online) MF (a) and QMC (b) results for 
$\langle S^zS^z \rangle _k$
in a homogeneous system with $N=102$, $N_f=66$, and $U=6t$.}
\label{spinesK102o066}
\end{figure}

(ii) At any noncommensurate filling, the Hubbard model in 1D
describes a metal (Luttinger liquid), so that no gap appears in the
charge and spin excitations, and there is a $2 k_F$ modulation in
the spin-spin correlation function that leads to the well-known
$2k_F$ singularity. Within the MF approximation, there is 
always a band insulator at any noncommensurate filling, with a gap that 
decreases when the system departs from half filling. The appearance of 
this gap for any density is due to the perfect nesting present in one 
dimension. The insulating nature of these solutions can be seen 
in the global compressibility of the system that is always zero, 
in the behavior of the density-density and the spin-spin correlations 
which decay exponentially, and in the momentum distribution function where 
there is no Fermi surface, as shown in Fig.\ \ref{perfilK102o066MF} for 
$N=102$, $N_f=66$, and $U=6t$.
Within this MF approach, there is a $2k_F$ modulation 
of the magnetization only in the $z$ component [the $SU(2)$ symmetry was 
broken], which leads to a divergence of the Fourier transform of 
$\langle S^zS^z\rangle$ at $k=2k_F$. In Fig.\ \ref{spinesK102o066} we 
compare the MF result (a) for $\langle S^zS^z \rangle _k$ with the 
QMC one (b), for $N=102$, $N_f=66$, and $U=6t$, where it can be seen that 
the $2k_F$ peak is one order of magnitude bigger in the MF case compared 
to the QMC case. This is due to the existence of the magnetization 
in the MF solution (the values at the peaks will only diverge in the 
thermodynamic limit).

\begin{figure}[h]
\includegraphics[width=0.48\textwidth,height=0.57\textwidth]
{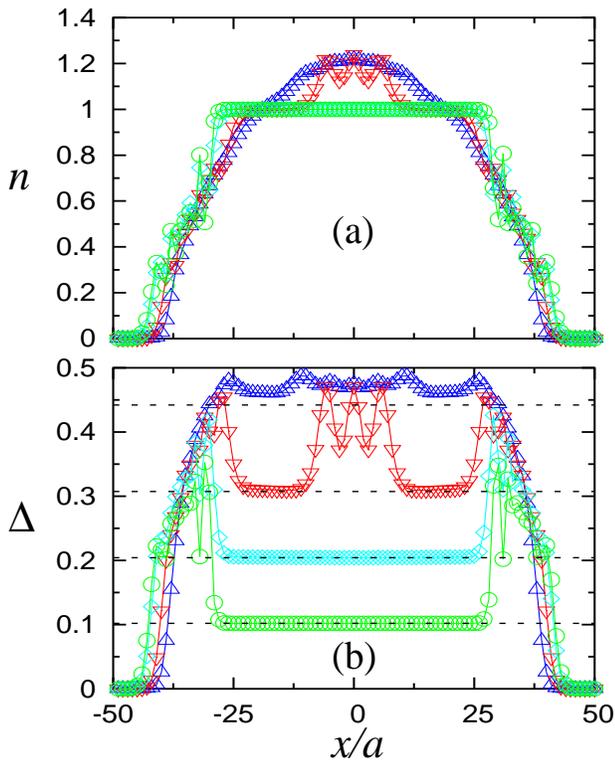}
\caption{(Color online) MF profiles for a trap with $N=100$, 
$Va^2=0.0025t$, and 
$N_f=70$, the on-site repulsions are 
$U/t=2$ (\textcolor{blue}{$\triangle$}), 
3 (\textcolor{red}{$\bigtriangledown$}),
4 (\textcolor{cyan}{\large $\Diamond$}),
and 6 (\textcolor{green}{$\bigcirc$}). (a) Local density,
(b) variance of the local density. 
The dashed lines in (b) are the MF values
of the variance in the $n=1$ homogeneous system for
$U/t=2, \ 3,\ 4,\ 6$ (from top to bottom).}
\label{perfdelt100MF}
\end{figure}

\begin{figure}[h]
\begin{center}
\includegraphics[width=0.77\textwidth,height=0.93\textwidth]
{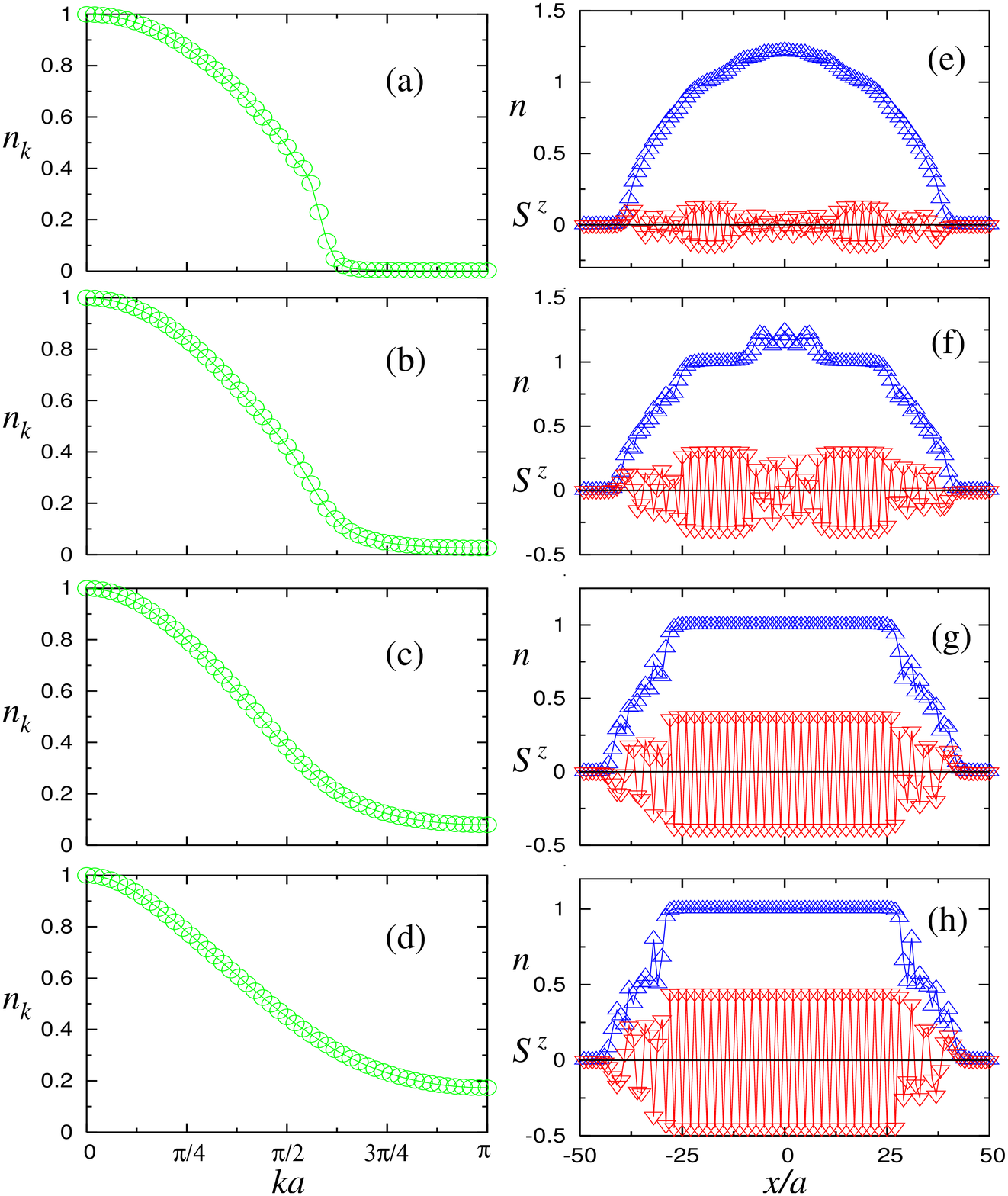}
\caption{(Color online) MF normalized momentum distribution function 
(\textcolor{green}{$\bigcirc$}) (a)-(d), 
and their corresponding density (\textcolor{blue}{$\triangle$}) and 
$\langle S^z \rangle$ (\textcolor{red}{$\bigtriangledown$}) 
profiles (e)-(h) for $U/t=2$ (a),(e), 3 (b),(f), 4 (c),(g), 6 (d),(h) 
and $N=100$, $N_f=70$, $Va^2=0.0025t$.}
\label{perfilK100o070MF}
\end{center}
\end{figure}

We consider next the confined inhomogeneous case. 
In Fig.\ \ref{perfdelt100MF}(a) we show the density profiles obtained
for a trap with $N=100$, $N_f=70$, and $Va^2=0.0025t$ (like the one
presented in Fig.\ \ref{perfdeltcomp100}) for different values of the 
on-site repulsion $U/t=2, \ 3, \ 4, \ 6$ [for more details see Figs.\ 
\ref{perfilK100o070MF}(e)-\ref{perfilK100o070MF}(h)].
It can be seen that the MF solutions have also density profiles in which 
``metallic'' ($n\neq 1$) and insulating ($n=1$) phases coexist. However, 
the MF insulating plateaus with $n=1$ [$U/t=4$ in Fig.\ 
\ref{perfdelt100MF}(a)] appear for smaller values of $U$ than the ones 
required in the Hubbard model for the formation of the Mott-insulating plateaus 
[$U/t=6$ in Fig.\ \ref{perfdeltcomp100}(a)]. In the ``metallic''
phases it is possible to see that the charge density shows rapid spatial 
variations that are very large for $n>1, U/t=4$ [see also Fig.\
\ref{perfilK100o070MF}(f)]. These density variations in the 
``metallic'' phases are reflected in the variance profiles [Fig.\ 
\ref{perfdelt100MF}(b)]. The values of the variance in 
the plateaus are the same as the ones obtained in the homogeneous MF 
case for the same value of $U$ when $n=1$ [Fig.\ \ref{perfdelt100MF} 
(b)]. This feature was discussed in Sec.\ II for the QMC solutions. 

\begin{figure}[h]
\includegraphics[width=0.44\textwidth,height=0.27\textwidth]
{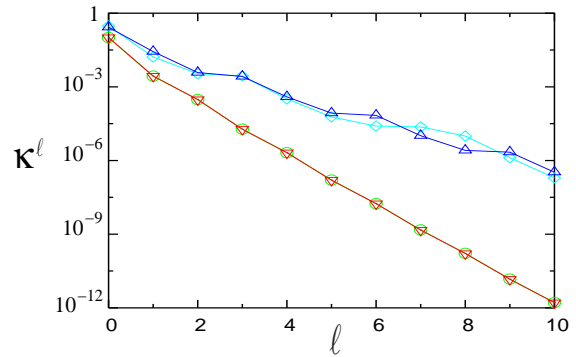}
\caption{(Color online) MF local compressibility as a function of 
$\ell$ for the 
homogeneous and trapped ($Va^2=0.0025t$, $N_f=70$) systems for $n=1$ 
and $n \sim 0.66$, $U=6t$. 
(\textcolor{red}{$\bigtriangledown$}) $n=1$ homogeneous,  
(\textcolor{green}{$\bigcirc$}) $n_{i=0}=1$ trapped, 
(\textcolor{blue}{$\triangle$}) $n \sim 0.66$ homogeneous, and
(\textcolor{cyan}{\large $\Diamond$}) $n_{i=-30}\sim 0.66$ trapped. }
\label{Compvsl_MF}
\end{figure}

Since the MF solution leads to an insulating state 
for incommensurate fillings in the homogeneous case, we analyze in
the following more carefully the regions with $n\ne 1$ in the presence of 
the trap. We find that the local compressibility 
always decays exponentially as a function of $\ell$ over the entire 
system, although the exponents are different depending on the density of 
the point analyzed. Some additional modulation appears in $\kappa^{\ell}$ 
when $n\ne 1$, as shown in Fig.\ \ref{Compvsl_MF} where the local 
compressibility is displayed as a function of $\ell$ for $n=1$ and 
$n\sim 0.66$ in the trapped and homogeneous cases. 
Therefore, although the MF approximation leads to a density profile 
similar to the one obtained in QMC as long as $n<1$, it gives a
qualitative wrong description of the character of those regions. 

The MF results for the normalized momentum distribution function are
presented in Figs.\ 
\ref{perfilK100o070MF}(a)-\ref{perfilK100o070MF}(d) with their 
corresponding density profiles [Figs.\ 
\ref{perfilK100o070MF}(e)-\ref{perfilK100o070MF}(h)], 
for the same parameter values as 
in Fig.\ \ref{perfdelt100MF}. It can be seen that the normalized momentum 
profiles are similar to the ones obtained with QMC when the density 
profiles are similar, so that the MF results are very similar to the QMC 
ones for this 
averaged quantity although the physical situation is very different. 
(Within MF there is always an insulator in the system and 
within QMC there are metallic and insulating phases coexisting.) 
In Figs.\ \ref{perfilK100o070MF}(e)-\ref{perfilK100o070MF}(h), we also show the 
$\langle S^z \rangle$ component of the spin on each site of the trap 
[$S^z=\left( \hat{n}_{i\uparrow }-\hat{n}_{i\downarrow }\right)/2$].
When the density is around $n=1$, it can be seen that
antiferromagnetic order appears, and that a different
modulation exists for $n \neq 1$. We should mention that we have also
found some MF solutions for the trapped system where there was formation 
of spin domain walls in the insulating regions with $n=1$ and there the 
variance was not constant (a plateau in the density does not correspond 
to a plateau in the variance), so that care should be taken with
the mean field solutions.

\onecolumngrid

\newpage\

\twocolumngrid

\section{Phase Diagram}

In the present section, we study the phase diagram for fermions 
confined in harmonic traps with an underlying lattice.

\begin{figure}[h]
\includegraphics[width=0.4\textwidth,height=0.29\textwidth]{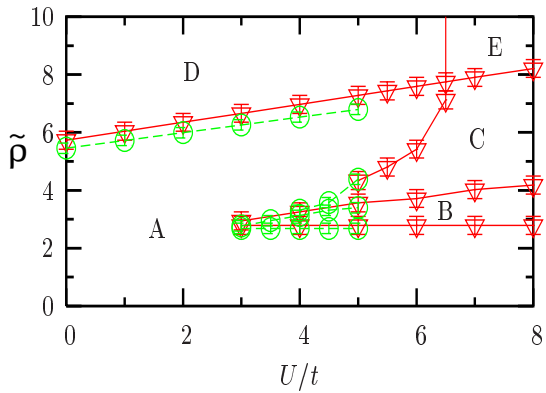}
\caption{(Color online) Phase diagram for systems with $Va^2=0.006 t, 
\ N=100$ 
(\textcolor{red}{$\bigtriangledown$}) and $Va^2=0.002t, \ N = 150$ 
(\textcolor{green}{$\bigcirc$}). The different phases are explained 
in the text.} \label{DFS_ScalingF}
\end{figure}

As it can be inferred from Fig.\ \ref{perfil3D}, the phases present 
in the system are very sensitive to the values of the parameters in 
the model, i.e., to the curvature of the parabolic potential
($V$), to the number of fermions present in the trap ($N_f$) and
to the strength of the on-site repulsion ($U$). In order to be able to 
relate systems with different values of these parameters and different 
sizes, we use the characteristic density ($\tilde{\rho}$)
defined in Sec.\ III for the definition of the phase diagram. 
This quantity has been shown to be a meaningful quantity in the 
thermodynamic limit since density profiles (as function of $x/\zeta$) 
and normalized momentum distributions do not change when $\tilde{\rho}$ 
is kept constant and the filling in the system (or the occupied 
system size) is increased. For a harmonic potential, 
the characteristic length is given by $\zeta=\left( V/t \right)^{-1/2}$ and 
the characteristic density is then $\tilde{\rho}=
N_f a \left( V/t \right)^{1/2}$. In Fig.\ \ref{DFS_ScalingF} we
show the phase diagrams of two systems with different strengths of
the confining potential ($Va^2=0.006t$ and $Va^2=0.002t$) and different
sizes ($N=100$ and $N=150$, respectively). This shows that the
characteristic density is a meaningful quantity to characterize a
generic phase diagram. It allows to compare different systems,
and, hence, to relate the results of numerical simulations with larger 
experimental sizes (although currently the linear dimension of the 
experiments is $\sim 65\ a$).

The phases present in the phase diagram in Fig.\
\ref{DFS_ScalingF} can be described as follows: (A) a pure metal without
insulating regions, (B) a Mott insulator in the middle of the trap
always surrounded by a metal, (C) a metallic region with $n>1$ in
the center of the Mott-insulating phase, that is surrounded by a
metal with $n<1$, (D) an insulator with $n=2$ in the middle of the
trap surrounded only by a metallic phase, and finally (E) a ``band
insulator'' in the middle of the trap
surrounded by metal with $n>1$, and these two phases inside a
Mott insulator that as always is surrounded by a metal with $n<1$.
As it can be seen, this phase diagram is more complex than the
one for the 1D homogeneous Hubbard model. 

\begin{figure}[h]
\includegraphics[width=0.47\textwidth,height=0.49\textwidth]
{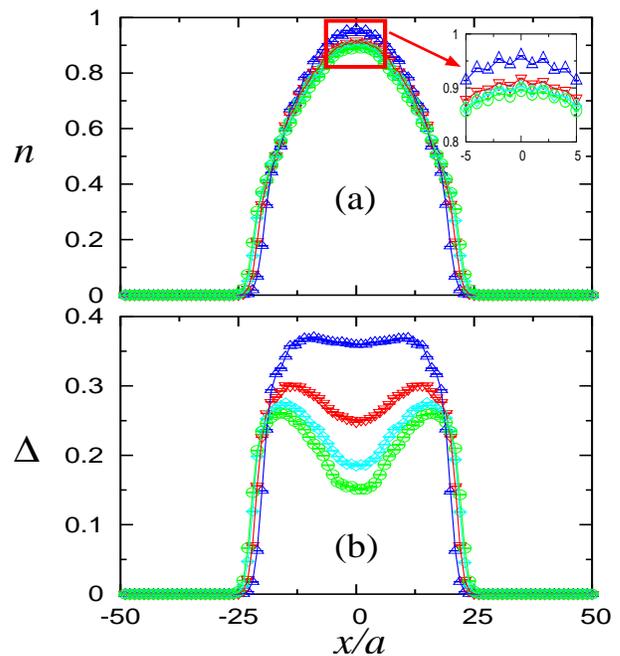}
\caption{(Color online) Profiles for a trap with $Va^2=0.006t$ 
and $N_f=30$, the 
on-site repulsions are $U/t=2$ (\textcolor{blue}{$\triangle$}), 
4 (\textcolor{red}{$\bigtriangledown$}), 
6 (\textcolor{cyan}{$\Diamond$}), 
and 8 (\textcolor{green}{$\bigcirc$}). 
(a) Local density, (b) variance of the local density.}
\label{Perfdelta100o030}
\end{figure}

There are some features that we find interesting in this 
phase diagram: (i) For all the values of $U$ that we have
simulated, we see that the lower boundary of phase B with the
phase A has a constant value of the characteristic density. (ii) We
have also seen that the upper boundary of phase B with phases A and C
and the boundary of phases D and E with A and C, respectively, are
linear within our errors. (iii) Finally, there is another
characteristic of phase B, and of the metallic phase A that is
below phase B, that we find intriguing and could be
related to point (i). As can be seen in Fig.\ \ref{perfdeltcomp100}(a) 
for $U/t=6$ and 8, once the Mott-insulating phase is formed in the 
middle of the trap, a further increase of the on-site repulsion leaves the 
density profile almost unchanged. (The two mentioned profiles are one 
on top of the other in that figure.) However, if we look at the
variance of these profiles it decreases with increasing $U$, which means
that the double occupancy 
$\left\langle n_\uparrow n_\downarrow\right\rangle  = 
\left[ \Delta - n \left(1 - n \right) \right]/2$ is decreasing in the system
when $U$ is increased, without a redistribution of the density. 
Less expected for us is that this also occurs for the
metallic phase A that is below phase B. In Fig.\
\ref{Perfdelta100o030}(a) we show four density profiles and 
in Fig.\ \ref{Perfdelta100o030}(b) their variances, for a trap
with $Va^2=0.006t$ and $N_f=30$ when the on-site repulsion has
values $U/t=2$, 4, 6, and 8. If we look at the phase diagram, the
point with $U/t=2$ is not below phase B whereas the other three
values are. In Fig.\ \ref{Perfdelta100o030}(a) we can see that as
the value of $U/t$ is increased from 2 to 4 there is a visible change in
the density profile, but after a further increase the profile
remains almost the same for the other values of $U$. The inset
shows a magnification of the top of the profiles for a more
detailed view. Fig.\ \ref{Perfdelta100o030}(b) shows that although 
the density stays almost constant for $U/t=4,\ 6$, and 8, the variance 
decreases. In the other phases (C, D, and E), an increase of the on-site 
repulsion always changes the local densities pushing particles to 
the edges of the trap.

\begin{figure}[h]
\includegraphics[width=0.46\textwidth,height=0.31\textwidth]
{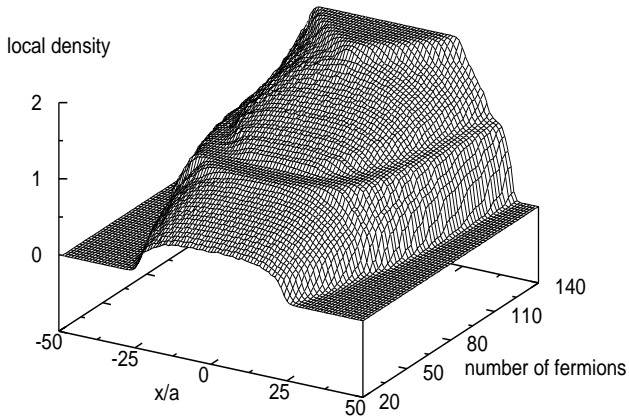}
\caption{Evolution of the local density in a quartic confining potential 
as a function of the position in the trap and increasing total number of
fermions. The parameters involved are $N=100$, $U=5t$, and 
$V_4a^4=3 \times 10^{-6}t$.}
\label{perfil3DV4}
\end{figure}

\begin{figure}[h]
\includegraphics[width=0.39\textwidth,height=0.28\textwidth]
{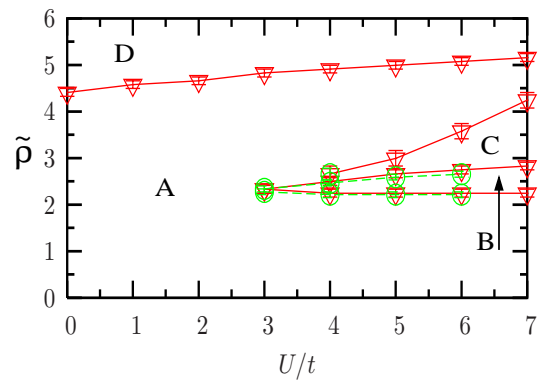}
\caption{(Color online) Phase diagram for systems with 
$V_4a^4=3 \times 10^{-6}t, 
\ N=100$ (\textcolor{red}{$\bigtriangledown$}) and 
$V_4a^4=10^{-6} t, \ N = 100$ (\textcolor{green}{$\bigcirc$}). 
The phases are labeled in the same way as in Fig.\ \ref{DFS_ScalingF}.}
\label{DFS_ScalingV4F}
\end{figure}

Finally, in order to see, whether the features of the phase diagram 
discussed previously are particular of a harmonic confining potential, 
we have also performed an analogous study 
for a quartic confining potential. In Fig.\ \ref{perfil3DV4} we show how 
the density profiles evolve in a trap with a quartic confining potential 
($V_4a^4=3 \times 10^{-6}t$) when the total filling is increased from 
$N_f=20$ to 140. It can be seen that the shape of the metallic regions is 
flatter than in the harmonic case. However, all the 
features discussed previously for the parabolic confining potential 
are also present in the quartic case. Local phases appear in the system, 
the Mott-insulating plateaus with $n=1$ have values of the variance 
equal to the ones in the homogeneous case for the same value of $U$, 
the local compressibility vanishes in the Mott-insulating regions and 
the spin and momentum distribution exhibit a behavior similar to that of 
the parabolic case. 
In addition, the form of the phase diagram for the 
quartic confining potential is similar to that of the harmonic trap. 
In Fig.\ \ref{DFS_ScalingV4F} we show the phase diagram for a system with 
$N=100$ and $V_4a^4=3 \times 10^{-6}t$, where the characteristic density 
is given by $\tilde{\rho}=N_f a\left( V_4/t \right)^{1/4}$. 
We have also calculated some points of the phase diagram for another 
system with $V_4a^4=10^{-6}t$ and $N=100$ in order to check that 
the scaling relation also works properly in this case. The phases are 
labeled in the same way as in Fig.\ \ref{DFS_ScalingF}, up to the largest 
value of $U$ that we have studied we do not obtain the phase E which for 
the quartic confining case seems to be moved further towards the strong 
coupling regime. Figure \ref{DFS_ScalingV4F} shows that the phase diagram 
in this case is very similar to that of the parabolic case, so that all 
the features discussed previously apply to the quartic confining 
potential. The form of the phase diagram in Fig.\ \ref{DFS_ScalingF} 
seems to be of general applicability for any confining potential when the 
proper characteristic density is considered. 

\section{Conclusions}

We have studied the ground-state properties of the 1D fermionic 
Hubbard model in harmonic traps with an underlying lattice, 
using QMC simulations and a MF approach. 

Due to the inhomogeneous density distribution, in general metallic
and insulating regions coexist in the trap. Therefore, local quantities
have to be used to characterize the system. 
We considered first the variance of the local density, which, when
a plateau develops in the density profile with $n=1$, equals the value
in the homogeneous system in the Mott-insulating phase. However,
it is not an unambiguous quantity, since it can be larger in the  
Mott-insulating phases than in metallic regions with densities close to
$n=0$ or $n=2$. Therefore, we have defined a
local-order parameter (local compressibility) that vanishes in
the insulating regions and is finite in the metallic ones.
The local compressibility gives a measure of the local change in 
density due to a constant shift of the 
potential over finite distances larger than the correlation
length of the density-density correlation function
for the Mott-insulating phase in the unconfined system. As expected, antiferromagnetic 
correlations are present in the Mott-insulating phases and they remain in 
the shoulders of the density profile for $n \sim 1$.

As opposed to homogeneous or periodic systems, where the appearance of
a gap can in general be seen by the disappearance of the Fermi edge in the
momentum distribution function, it is easily seen that due to the presence 
of a confining potential, such a feature is much less evident. 
Although increasing the power of the confining potential 
sharpens the features connected with the Fermi edge, 
it remains qualitatively different from the homogeneous case.
Increasing the system size with a proper definition of the
thermodynamic limit, does 
not change at all the behavior of the density and momentum distribution 
functions in the trapped case, such that, in fermionic systems it does not 
seem to be the appropriate quantity to look for evidence of gaps opening
in the system.

The MF study has shown that within this approach the trapped system is 
an insulator for all the densities in the trap, a qualitatively
wrong picture in contradiction with the QMC results. In addition, 
the metallic regions showed large oscillations in the density that 
are not present in the real solutions. However, it is 
possible to obtain some qualitative information from MF, such as the 
coexistence of $n\ne 1$ regions with $n=1$ plateaus, the shape of the 
momentum distribution function and the existence of antiferromagnetic 
correlations in the insulating regions with $n=1$.

Finally, we have determined a generic form for the phase diagram that
allows to compare systems with different values of all the
parameters involved in the model. It can be used to predict the phases that 
will be present in future experimental results. The phase diagram also 
reveals interesting features such as reentrant behavior
in some phases when some parameters are changed and phase boundaries 
with linear forms. A similar form of the phase 
diagram has been also found for a quartic confining potential, 
such that the results obtained here are 
generally applicable to cases beyond a perfect harmonic potential.

\begin{acknowledgments}

We gratefully acknowledge financial support from the LFSP 
Nanomaterialien. We are grateful to G. G. Batrouni and R. T. Scalettar
for interesting discussions at the early stages of the project. We are 
grateful to F. F. Assaad for valuable discussions about the MF approach 
we used here and also for providing us the core of the code we employed 
in our MF calculations. We thank R. Noack and G. Modugno for useful 
comments on the manuscript, and M. Feldbacher and T. Pfau for useful 
conversations. We thank HLR-Stuttgart (Project DynMet) for allocation of 
computer time, the calculations were carried out on the HITACHI SR8000.

\end{acknowledgments}

\end{document}